\documentclass[12pt,a4paper,twoside]{paper}
\usepackage{amssymb}
\usepackage{graphicx}
\usepackage{epstopdf}
\usepackage{amsmath}

\newcommand{\be}{\begin{equation}}
\newcommand{\ee}{\end{equation}}

\newcommand{\ba}{\begin{eqnarray}}
\newcommand{\ea}{\end{eqnarray}}

\newcommand{\bea}{\begin{eqnarray*}}
\newcommand{\eea}{\end{eqnarray*}}

\newcommand{\bee}{\begin{enumerate}}
\newcommand{\ene}{\end{enumerate}}

\usepackage{epsfig}
\def\R{\mathbb R}

\def\Z{\mathbb Z}

\begin{document}

     \vspace*{7mm}

     \vspace*{10mm}

     \begin{center}
 \textsl{\Huge    Probabilistic Explanations and the Derivation of Macroscopic Laws\footnote{Invited paper for a volume on the philosophy of statistical mechanics, titled ``Statistical Mechanics and Scientific Explanation: Determinism, Indeterminism, and the Laws of Nature," edited by Valia Allori.}
}

 \vspace*{10mm}

 { \Huge Jean Bricmont\footnote{ 
 IRMP,
Universit\'e catholique de Louvain,
chemin du Cyclotron 2,
1348 Louvain-la-Neuve,
Belgium. E-mail: jean.bricmont@uclouvain.be}
}
\end{center}

\begin{abstract}

{ \Large 
We will discuss the link between scientific explanations and probabilities, specially in relationship with statistical mechanics
 and the derivation of macroscopic laws from microscopic ones.
 }
\end{abstract}

\newpage

\vspace*{5mm}

\section{Introduction}\label{sec1}

It is a commonplace that macroscopic laws, in particular the second law of thermodynamics (``entropy increases"), are true only in a probabilistic sense. But if one restricts oneself to classical physics, its laws are deterministic, so one might ask: where do these probabilities come from? How to make sense of objective probabilities in a deterministic universe? And if those probabilities 
 are in some sense ``subjective", namely assigned by us to events, and not ``intrinsic" to those events, how can one say that macroscopic laws are objective?  

Since the flow of heat from hot to cold is a perfectly objective fact and is law-like in its universality, should one say that our explanations of this fact are unsatisfactory if they rely on non-objective probabilities?

Our goal here is to try to answer these questions and to disentangle certain confusions that those questions  tend to create.

We will start with general considerations about objectivity and subjectivity in science and also in different  notions of probability.

Then we will explain how probabilities enter in the  explanation of both equilibrium statistical mechanics and in the approach to equilibrium. We will illustrate the latter through a simple example, the Kac ring model. For related work in the same spirit, see e.g. \cite{BCV, CCCV,  dBP}.

\section{Objectivity and Subjectivity }\label{sec2}

There is a constant tension in the history of philosophy (and of science) between people who think that our thoughts are produced by 
our minds with little connection to the world (`idealists') and those who  think that they are the result of an interaction between our mind and a mind-independent ``outside world" (`realists'). At one extreme, one finds solipsism, everything going on in our mind is just like a dream or an internal movie, at the other extreme, one finds naive realists,  for whom reality is ``objectively" what it looks like, including colors, odors etc. 
It is not my ambition here to resolve this issue or even to discuss it adequately, although I am on the (non-naive) realist side.

Coming back to science, one may distinguish, even from a realist point of view, different degrees of ``objectivity":

\begin{itemize}
\item[1.] {\it Facts.} It is about facts that our intuition of objectivity is strongest: I am right now writing on a computer, the moon is there even if nobody looks at it, rivers flows and the sun shines when it does. An idealist might deny all this, but it is hard to be a realist without accepting the existence of facts ``out there", independent of my consciousness. 

\item[2.] {\it Laws.} We all know that there are regularities in Nature and it is a priori reasonable to think of those regularities as being part of Nature and to call them laws of Nature. Someone who is a realist about facts might deny the reality of laws over and above the so-called Humean mosaic of empirical facts, considering them as a mere human tool to summarize the observed regularities.\footnote{See Maudlin \cite{Ma} for a  discussion of realism about laws versus the ``Humean" conception.}  We will not analyze this issue here  and simply admit, for the sake of the discussion, the objectivity of the laws of Nature.
Moreover, one can introduce a hierarchy of such laws: there are fundamental laws, governing the behavior of the most microscopic constituents of matter and derived or phenomenological laws describing the behavior of aggregate sets of such particles. For example, if ``heat is molecular motion", then the laws governing heat are phenomenological but those governing the molecular motions are fundamental.\footnote{At least to a first approximation: one may consider the molecules as being made of more fundamental entities, whose laws would be truly fundamental, while those governing the molecules would be phenomenological. But, since we do not know what are really the ultimate constituent of matter, this fundamental/phenomenological distinction is relative to a given context.}

\item[3.] {\it Explanations.} There is a common (mis)-conception according to which the role of science is to describe and to predict but not to explain. However, if one asks: "why does it rain today?", one is asking for an explanation of a given fact and the answer will involve laws of meteorology and empirical data concerning that past situation of the atmosphere. So, it will be an explanation and it will be scientific.

 Science is in fact full of explanations: the theory of gravitation explains  the regularities in the motion of planets, moons or satellites. The atomic theory of matter explains the proportions of elements in chemical reactions. Medical science explains in principle what cures a disease, etc.

The misconception arises because people often think of ``ultimate" explanations, like: why is there something rather than nothing? Or, how did the Universe (including the Big Bang) come to exist? But, if one puts aside those metaphysical/religious questions to which nobody has an answer, science does provide explanations of observable phenomena. 

Of course, the question:  ``what constitutes a valid explanation?", specially when probabilities are involved, is a tricky one and we will discuss it in sections \ref{sec5} and \ref{sec6}.

\item[4.] {\it Probabilities.} 
If one puts oneself in the framework of classical physics (in order to avoid quantum subtleties) then laws are deterministic which means that, given some initial conditions, future events either occur or do not occur. There is no sense in which they are, by themselves,  probable or improbable. Yet, scientists use probabilities all the time. Although there is a school of thought that tries to give an objective meaning to the  notion of probability (we will discuss it below), there must be something ``human" about probabilities in the sense that we may use them in certain ways and have good reasons to do so, but probabilities are not expressing objective facts about the world,  independent of us, like, say, the motion of the moon.

 \end{itemize}

\section{Two Notions of Probability }\label{sec3} 
 
 There are, traditionally, at least two different meanings given to the word `probability'
in the natural sciences. These two meanings are well-known but, since much
confusion arises from the fact that the same word is used to denote two very different
concepts, let us start by recalling them and explain how one can connect
the two. First, we speak of  `the natural sciences', because we  do not want to discuss
the purely mathematical notion of probability as measure (the notion introduced by
Kolmogorov and others). This is of course an important branch of mathematics and
the notion of probability used here, when formalized, will coincide with that
mathematical notion, but we want to focus here on the role played by probabilities
in our scientific theories, which is not reducible to a purely mathematical concept.

So, the first notion that comes to mind is the so-called `objective' or `statistical'
one, i.e. the view of probability as something like a `theoretical frequency': if one
says that the probability of the event E under condition X, Y, Z equals p, one means
that, if one reproduces the conditions X, Y, Z sufficiently often, the event E will appear
with frequency p. Of course, `sufficiently often' is vague and this is the source of
much criticism of that notion of probability. But, putting that objection aside for
a moment and assuming that `sufficiently often' can be given a precise meaning in
concrete circumstances, probabilistic statements are, according to this view, factual
statements that can be confirmed or refuted by observations or experiments.

By contrast, the `subjective' or Bayesian use of the word `probability' refers to
a form of reasoning and not to a factual statement. Used
in that sense, assigning a probability to an event expresses a  rational judgment
on the likelihood of that single event, based on the information available at that
moment. Note that, here, one is not interested in what happens when one reproduces
many times the `same' event, as in the objective approach, but in the probability of
a single event. This is of course very important in practice: when I wonder whether
I need to take my umbrella because it will rain, or whether the stock market will
crash next week, I am not mainly interested in the frequencies with which such
events occur but with what will happen here and now; of course, these frequencies
may be part of the information that is used in arriving at a rational judgment on the probability of
a single event, but, in general, 
they are not  the only information available.

How does one assign subjective probabilities to an event? In elementary textbooks,
a probability is defined as the ratio between the number of favorable outcomes
and the number of `possible' ones. While the notion of favorable outcome is easy to
define, the one of possible outcome is much harder. Indeed, for a Laplacian demon,
nothing is uncertain and the only possible outcome is the actual one; hence, all probabilities are zeroes or ones. But we are not Laplacian demons and it is here that
ignorance enters.\footnote{This was of course Laplace's main point in \cite{La}, although this is is often misunderstood.
Laplace emphasized that human intelligence will forever remain `infinitely distant' from the one
of his demon.} We try to reduce ourselves to a series of cases about which we are
`equally ignorant', i.e. the information that we do have does not allow us to favour
one case over the other, and that defines the number of `possible' outcomes. The
standard examples include the throwing of a dice or of a coin, where the counting
is easy, but that situation is not typical.
At the time of Laplace, this method was called the `principle of indifference';
its modern version is the {\it maximum entropy principle}. Here one assigns to each
probability distribution ${\bf p}= (p_i)_{i=1}^N$  its Shannon entropy, given by:
$$
S({\bf p})= -\sum_{i=1}^N p_i \ln p_i.
$$ 
One then chooses the probability distribution that has the maximum entropy, among
those that satisfy certain constraints that incorporate the information that we have
about the system.

The rationale, like for the indifference principle, is not to introduce bias in our
judgments, namely information that we do not have (like people who believe in lucky
numbers). And one can reasonably argue that maximizing the Shannon entropy is
indeed the best way to formalize that notion, see \cite{Sh, Ja1}, \cite[section 11.3]{Ja2}.

In practice, one starts by identifying a space of states in which the system under
consideration can find itself and one assigns a prior distribution to it (maximizing
the Shannon entropy, given the information available at the initial time), which is then updated when new information becomes available.\footnote{For an introduction to
Bayesian updating, see e.g. \cite{Ja1, Ja2}.}
Note that probabilistic statements, understood subjectively, are forms of reasoning,
although not deductive ones. Therefore, one cannot check them empirically, because reasonings, whether they are
inductive or deductive, are either correct or not, but that depends on the nature of the reasoning not on any facts.

If someones says: Socrates is an angel; all angels are immortal; therefore Socrates
is immortal, it is a valid (deductive) reasoning. Likewise, if  all we know
about a coin is that it has two faces and that it looks symmetric, therefore the probability
of `head' is one half, it is a valid probabilistic reasoning; throwing the coin a
thousand times with a result that is always tails does not disprove the reasoning; it
only indicates that the initial assumption (of symmetry) was probably wrong (just
as watching Socrates dead leads one to reconsider the notion that he is an angel or
that the latter are immortal); the main point of Bayesianism is to give rules that
allow to update one's probabilistic estimates, given previous observations.

Let us now consider some frequent objections to this ``subjective" notion of probability.
 \begin{itemize}
 
\item[1.]  {\bf Subjectivism.} Some people think that a Bayesian view of
probabilities presupposes of some form of subjectivism, meant as a doctrine
in philosophy or philosophy of science that regards what we call
knowledge as basically produced by ``subjects" independently of any connection to the ``outside world".
But there is no logical link here: a subjectivist about probabilities may
very well claim that there are objective facts in the world and that the laws governing
it are also objective, and consider probabilities as being a tool used in situations where our
knowledge of those facts and those laws is incomplete. In fact, one could
argue that, if there is any connection between Bayesianism and philosophical
subjectivism, it goes in the opposite direction; a Bayesian should naturally
think that one and only one among the `possible' states is actually realized,
and that there is a difference between what really happens in the world and
what we know about it. But the philosophical subjectivist position often starts
by confusing the world and our knowledge of it (for example, much of loose talk
about everything being `information' often ignores the fact that `information' is
ultimately information about something which itself is not information). 

Besides, ignorance does enter in the computations of probabilities but, as we
will see in the next section, when we discuss the connection between probabilities
and physics, this does not mean that either knowledge or ignorance are
assumed to play a fundamental role in physics.

\item[2.]   {\bf Determinism.} One may object that  Bayesians are committed to 
 a deterministic view of the world: since  Bayesians regard probabilities as subjective, doesn't this deny the possibility that
phenomena be intrinsically or genuinely random?
 Not necessarily. A Bayesian may be agnostic concerning the issue of intrinsic randomness and 
point out that it is difficult to find an argument showing the presence of intrinsic
randomness in nature; indeed, it is well-known that some deterministic
dynamical systems (the `chaotic' ones) pass all the statistical tests that might
indicate the presence of `randomness'.\footnote{Here is a simple example of such a system. Let $I=[0,1[$ and let $f: I \to I$
be given by $f(x)= 2 x \mod 1$. Then, writing $x \in I$  as $x=\sum_{n=1}^\infty \frac {a_n}{2^n}$, with $a_n= 0, 1$, we see that the map $f$ is equivalent to the shift $\sigma$ on sequences ${\bf a}= (a_n)_{n=1}^\infty$,
$\sigma ({\bf a})_n= a_{n+1}$. Using this observation, and the fact that the Lebesgue measure on $I$ is equivalent to the product measure on the sequences ${\bf a}$ giving a weight $\frac {1}{2}$ to both $0$ and $1$, one can check that the map $f$ is equivalent to a sequence of ``random" coin tossings with $a_n=0$ being, say, ``head" and $a_n=1$ being ``tail". So, that simple deterministic system will look as random as any apparently random system. For more fancy ``chaotic" dynamical systems, see
  \cite{Bow, BR, ER, Sin, Ru1, Ru2}.
} 
So, how can we know, when we observe
some irregular and unpredictable phenomenon, that this phenomenon is
`intrinsically random' rather than simply governed by unknown, but `chaotic', deterministic
laws?

\item[3.]   {\bf (Ir)relevance to physics.} One may think that the Bayesian approach is useful in games of chance or in
various practical problems of forecasting (like in insurances) but not for physics. Our answer will be based on the law of large numbers discussed in the next section.

   \end{itemize} 

The main point of this discussion is that there is nothing arbitrary or subjective in the assignment of ``subjective" probabilities.
The word ``subjective" here  simply refers to the fact that there are no true or real probabilities ``out there". But the choice of probabilities obeys rules (maximizing Shannon's entropy and doing Bayesian updating) that do not depend  of any individual's whims, although it does depend on his or her information.\footnote{A further confusion arrises from the fact that some probabilists, the Italian Bruno de Finetti being the best known one, do consider probabilities as expressing purely subjective degrees of beliefs that are constrained only by rules of consistency, see e.g. \cite{dF1,dF2}. These probabilists are sometimes called ``subjective Bayesians"; the view presented here is then called ``objective Bayesian", see \cite[p. 4]{Ja1} and \cite[p. 655]{Ja2} for a discussion of the difference between these views.}

  \section{The Law of Large Numbers }\label{sec4}
   
   A way to make a connection between the two views on probability goes through
the law of large numbers: the calculus of probabilities -- viewed now as part of
deductive reasoning -- leads one to ascribe subjective probabilities close to one for
certain events that are precisely those that the objective approach deals with, namely
the frequencies with which some events occur, when we repeat many times the
`same' experiment. So, rather than opposing the two views, one should carefully
distinguish them, but regard the objective one as, in a sense, derived from the
subjective one (i.e. when the law of large numbers leads to subjective probabilities
sufficiently close to one).
Let us state the law of large numbers, using a terminology that will be
useful when we turn to statistical mechanics below. Consider the simple example of
coin tossing. Let $0$ denote `head' and $1$, `tail'. The `space' of results of any single
tossing, $\{0, 1\}$, will be called the `individual phase space' while the space of all possible results
of $N$ tossings, 
 $\{0, 1\}^N$, will be called the `total phase space'. In statistical physics, the individual phase space will be $\R^3$
(if one considers only the positions or only the velocities of the particles) or $\R^6$ (if one considers both the positions and the velocities 
of the particles) and the total phase space will be $\R^{3N}$ or $\R^{6N}$ for $N$ particles.
   The variables $N_0$, $N_1$ that count
the number of heads $(0)$ or tails $(1)$ in $N$ tossings are  called {\it macroscopic}. 
   
   Here we introduce
an essential distinction between the {\it macroscopic variables}, or the {\it macrostate}, and
the {\it microstate}. The microstate, for $N$ tossing, is the sequence of results for all the
tossings, while the macrostate simply specifies the values of $N_0$ and $N_1$. Although this
example is trivial, let us draw the following analogy with statistical mechanics: $N_0$ and $N_1$ for a given point in the total
 phase space (a sequence of results for all the
tossings), count the number of `particles' that belong to a given subset ($0$ or $1$)
of the individual phase space. 

Now, fix $\epsilon > 0$ and define a sequence of sets of microstates ${\cal T_N} \subset \{0, 1\}^N$ to be {\it typical}, for  a given sequence of probability measures $P_N$ on $ \{0, 1\}^N$,
if 
\be
P_N({\cal T_N}) \to 1.
 \label{typ}
\ee
as $N \to \infty$.
If the typical sets ${\cal T_N}$ are defined by a property, we will also call that property typical.\footnote{This use of the word typical is not exactly the usual one, which refers to the probability of a given set, not a sequence of sets, to be close to $1$.}

Let $G_N(\epsilon
)$ be the set of  microstates such that 
 \be
 |\frac{N_0}{N}-\frac{1}{2}| \leq \epsilon
 \label{G}
\ee
 Here the letter $G$ stand for ``good", because we will use the same expression below in the context of statistical mechanics.
 
 Then, (a weak form of) the law of large numbers
states that, $\forall \epsilon
>0$,  
\be
P_N(G_N(\epsilon
)) \to 1
 \label{typ1}
\ee
 as $N\to \infty$, where  $P_N$ the product measure on  $ \{0, 1\}^N$ that assigns independent probabilities $\frac{1}{2}$
to each outcome of each tossing. This is the measure that one would assign on the basis of the indifference principle: 
give an equal probability to all possible sequences of results. In other words, what (\ref{typ1}) expresses is that the sequence of sets $G_N(\epsilon
)$ is typical in the sense of definition (\ref{typ}), $\forall \epsilon
>0$.
 
  A more intuitive way to say the same thing is that, if
we simply count the number of microstates that belong to $G_N(\epsilon
)$, we find that they form
a fraction of the total number of microstates close to 1, for N large.

The situation becomes more complicated but more interesting if one tries to understand what could be a {\it probabilistic explanation}, like the explanation of the second law of thermodynamics.

\section{Explanations and probabilistic explanations }\label{sec5}

A first
form of scientific explanation is given by laws. If state A produces state B, according to deterministic laws, then
the occurrence of B can be explained by the occurrence of A and the existence of those laws.\footnote{This is the main idea behind the deductive nomological model, according to which scientific explanations are deductive arguments with laws as one of the premises (see Hempel \cite{He1}, Hempel and Oppenheim \cite{He2}).}
If A is prepared in
the laboratory, this kind of explanation is rather satisfactory, since the initial state A is produced by us.

But if B is
some natural phenomena, like today's weather and A is some meteorological condition yesterday,  then A itself has to be explained, and that leads 
potentially to an ``infinite" regress, going back in principle to the beginning of the universe. In practice, nobody goes back that far, and A is simply taken to be ``given", namely our explanations are in practice limited.

It is worth noting that there is something ``anthropomorphic" even in this type of explanation: for example if A is something very special, one will try to explain A as being caused by anterior events that are not so special. Otherwise our explanation of B in terms of A will look unsatisfactory. Both the situations A and B and the laws are perfectly objective but the notion of explanation is ``subjective" in the sense that it depends on what we,  humans, regard as a valid explanation. 

Consider now a situation where probabilities are involved, take the simplest example, coin tossing, and try to use that example to build up our intuition about what constitutes a valid explanation.

 First observe  that, if we toss a coin many times and we find
approximately half heads and half tails, we do not feel that there is anything special
to be explained. If, however, the result deviates strongly from that average, we'll
look for an explanation (e.g. by saying that the coin is biased). 
 
This leads to the following suggestion: suppose that we want to explain some
phenomenon when our knowledge of the past is such that this phenomenon could
not have been predicted with certainty (for coin tossing, the past would be the initial conditions of the coins when they are tossed). We will say that our knowledge, although
partial, is {\it sufficient} to explain that phenomenon if we would have predicted it using
Bayesian probabilities  and the information we had about the past. That notion
of `explanation' incorporates, of course, as a special case, the notion of explanation
based on laws. Also, it fits with our intuition concerning the coin-tossing situation
discussed above: being ignorant of any properties of the coin leads us to predict a
fraction of heads or tails around one-half. Hence, such a result is not surprising or,
in other words, does not ``need to be explained", while a deviation from it requires
an explanation.

Turning to physics, consider for example the Maxwellian distribution of velocities,
for a free gas of N particles of mass m. Let $\Delta (\vec u)$, be the cubic cell of size $\delta^3$ centered
around $\vec u \in (\delta \Z)^3$. Let ${\bf v}= (\vec v_1, \dots, \vec v_N) \in \R^{3N} $, with each $\vec v_i \in \R^3$,
be an element of the `phase space' of
the system (where the spatial coordinates are ignored), i.e. a configuration
of velocities for all the particles, which is what we call a {\it microstate} of the system. 

Define the {\it macrostate} by the set of variables $\{N_{\vec u} ({\bf v})\}_{\vec u\in (\delta \Z)^3}$:
\be
N_{\vec u}({\bf v})= |\{i | \vec v_i \in \Delta (\vec u),\; i \in \{ 1, \dots, N\} \}|.
\label{macro}
\ee
$N_{\vec u}(\bf v)$ is also called {\it the empirical distribution} corresponding to the phase space point $\bf v$.
It counts, for each $\Delta (\vec u)$ and for a given set of velocities of all the particles, the number of particles whose velocities lie
in  $\Delta (\vec u)$.

This is analogous to counting the number of heads or tails in a given sequence of coin tosses or the number of times a dice falls on a given face when it is thrown many times.

Let $G_N(\epsilon, \delta)$, for given $\epsilon, \delta$, be  the set of ``good"  vectors $\bf v$ for which
\be
|\frac {N_{\vec u}(\bf v)}{N} - \frac { \exp (- \frac {m |{\vec u}|^2}{2kT})}{(2\pi mkT)^{3/2}}|\leq \epsilon,
\label{Maxw}
\ee
$\forall \vec u \in (\delta \Z)^3$.

Let  $S_{E,N}$ be the constant energy surface of energy E, namely the subset of $\R^{3N} $ defined by: 
\be
S_{E,N} = \{ {\bf v}= (\vec v_1, \dots, \vec v_N) | \sum_{i=1}^N  \frac { m |\vec v_i|^2}{2} =E \}
\label{S_E}
\ee
and let $\mu_{E,N}$ be the uniform measure on that surface (i.e. the restriction of the Lebesgue measure  in $\R^{3N} $ to that surface). 

Then, a variant of the
law of large numbers says that, for every  $\epsilon, \delta$,  the sequence of sets $G_N(\epsilon, \delta)$ is {\it typical}, in the sense of (\ref{typ}) for the sequence of probability measures 
 $\mu_{E,N}$, if $T$ in (\ref{Maxw}) is related to $E$ in (\ref{S_E}) by $kT = \frac { 2E}{ 3N}$, which means that  the set $G_N(\epsilon, \delta)$ 
 has, for N
large, a measure $\mu_{E,N}$ close to one when $kT = \frac { 2E}{ 3N}$ holds.
This is a precise way of saying that the distribution of velocities for a gas of N particles of mass m is  Maxwellian.

If someone asks: how does one explain the occurrence of this Maxwellian distribution?
The Bayesian answer is basically that there is nothing to explain, because this
is analogous to the situation of coin tossings when the fractions of heads and tails
are both close to one half. Given that we know that the energy is conserved, symmetry
considerations show that the uniform measure is the most natural one and, since the
Maxwellian distribution is the empirical distribution corresponding to most phase
points (relative to that measure), it is exactly what we would expect if we know
nothing more about the system. In fact, the only thing that would lead us {\it not} to
predict the Maxwellian distribution would be some additional knowledge about the
system (e.g. that there are some constraints or some external forces acting on it).

This answers the often heard question: ``how does one justify the choice of the
equilibrium measure?", namely here the measure $\mu_{E,N}$ of $S_{E,N}$: it is
the natural choice on Bayesian grounds. 

However, one can ask a related question, which is less trivial: how does one
explain the approach to equilibrium for a closed system which starts in a nonequilibrium configuration? This is the question that Boltzmann's analysis
answers and is the topic of the next section.

\section{Time evolution and probabilistic explanations }\label{sec6}

\subsection{Microstates and macrostates}\label{sec6.1}

We start by generalizing the notion of macrostate introduced in section \ref{sec5}.
Let ${\bf x}(t)$ be the {\it microstate} of a classical mechanical system on $N$ particles, namely 
$${\bf x}(t)= ({\vec q_1}(t), {\vec q_2}(t), \dots {\vec q_N}(t), {\vec p_1}(t),{\vec p_2}(t), \dots, {\vec p_N}(t)) \in \R^{6N},
$$
 where $\vec q_i(t) \in \R^3$ and $\vec p_i(t) \in \R^3$ are the position and the momentum of the $i^{th}$ particle at time $t$.

Let  
$$
 {\bf x}(0)  \to  {\bf x}(t) = T^t {\bf x}(0)$$
   denote the flow in $R^{6N}$  induced by  Hamilton's equations for that system and let $\Omega$ denote
a bounded subset of $R^{6N}$ invariant under that flow (for example a bounded constant energy surface).

Here, a {\it macrostate} is simply a map $F: \Omega \to R^{n} $ with $n$, the number of macroscopic variables, being much smaller than $6N$: $n<< 6N$. 

We can give, using (\ref{macro}), a   simple example of such a map, by letting
 $F=F({\bf x})= (\frac{N_{\vec u}({\bf v})}{N})_{\vec u \in (\delta \Z)^3}$.\footnote{A small caveat: here the number $n$ of macroscopic variables seems to be infinite, since
$\vec u \in (\delta \Z)^3$, but the number of non-zero values of $N_{\vec u}({\bf v})$ is finite, since,  because of
 (\ref{S_E}), $N_{\vec u}({\bf v})$ will be $0$ if $|\vec u| >\frac{2E}{m}+ \frac{3\delta}{2} $.} To give another example, let, as above,
$\Delta (\vec u)$ be the cubic cell of size $\delta^3$ centered
around $\vec u \in (\delta \Z)^3$ and let ${\bf q}= (\vec q_1, \dots, \vec q_N) \in \R^{3N} $, 
be an element of the `configuration space' of
the system, i.e. a configuration
of the positions for all the particles.
Define
\be
N_{\vec u}({\bf q})= |\{i | \vec q_i \in \Delta (\vec u),\; i \in \{ 1, \dots, N\} \}|.
\label{macro1}
\ee
Assume that the particles are enclosed in a box $\Lambda$ which is a union of cubic cells of size $\delta^3$, and let $F=F({\bf x})= (\frac{N_{\vec u}({\bf q})}{N})_{\Delta (\vec u)\subset \Lambda}$. Here $n$ is the number of cubic cells of size $\delta^3$
 in $\Lambda$, or $\frac{|\Lambda|}{\delta^3}$.

Note that the density function is simply a continuous approximation to the function $F$ (obtained in the limit $N\to \infty$, $\delta \to 0$).
One could also do that in the space $\R^6$,
combining both  
positions and momenta of the particles. In that case, the continuous approximation to $F$ is Boltzmann's $f$ function.  

 Now, one can associate to the evolution $ {\bf x}(0)  \to  {\bf x}(t) = T^t {\bf x}(0)$ an {\it induced evolution} $ F_0 \to    F_t $, obtained by:
 \be
 F_0= F({\bf x}(0)) \to    F_t = F({\bf x}(t)).
 \label{F}
 \ee
 
A natural question is whether 
 the evolution of $F$ is
{\it autonomous}, i.e. independent of the ${\bf x}(0)$ mapped onto $F_0$. If it is, then one can say that the evolution of $F_t$, which is called a {\it macroscopic law},  has been {\it reduced to} or {\it derived from} the microscopic one $ {\bf x}(0)  \to  {\bf x}(t) $, in a straightforward way.

But such an autonomous evolution is, in general, impossible, because the evolution
${\bf x}(0)  \to  {\bf x}(t)$ is    {\it reversible} meaning that, if $I$ denotes the operation: 
\be
I({\bf x}(t))= ({\bf q_1}(t), {\bf q_2}(t), \dots {\bf q_N}(t), {-\bf p_1}(t),{-\bf p_2}(t), \dots, {-\bf p_N}(t)),
\label{I}
\ee
 one has:
\be
T^tIT^t {\bf x}(0)= I  {\bf x}(0),
\label{rev}
\ee
or, in words, if one lets the system evolve according to the dynamical laws for an amount of time $t$,  
if one then reverses the velocities (or the momenta), and if, finally, one  lets the system evolve according those same laws 
for the same amount of  time $t$, one gets the initial state with the velocities reversed.

But the evolution
$F_0 \to F_t$ is often {\it irreversible}, for example if $F$ is the density and  if one starts with a non-uniform density,
the evolution of $F$ tends to a uniform density and will not return to a non-uniform one.

Yet the reversibility argument shows that, since changing the sign of the velocities does not change the density, for
each microstate ${\bf x}(t)= T^t {\bf x}(0)$ giving  rise to a given 
value of $F_t= F({\bf x}(t))$, there may exist another microstate $I({\bf x}(t))$  giving  rise to the same 
value of $F_t= F(I({\bf x}(t)))$ but such that the future time evolution of $F_t$ will be markedly different 
depending on whether it is induced by ${\bf x}(t)$ or by $I({\bf x}(t))$. So, the evolution of the macrostate cannot be autonomous in the sense given here.

This seems to imply that one cannot derive a macroscopic law from a microscopic one and in particular that one cannot 
give a microscopic derivation of the second law of thermodynamics implying that the entropy monotonically increases.
Yet, as we will explain now, this can be done, but not in the straightforward way suggested above.

\subsection{Derivation of macroscopic laws from microscopic ones}\label{sec6.2}

The basis of the solution to the apparent difficulty mentioned in the previous subsection is that 
 {\it the map $F$ is many to one in a way that depends on value taken by $F$}!
 
 To explain that, think again of the simple example of $N$ coin tossings with $F$= number of heads and the microstates being
 the sequence of results e.g. $(H, T, H, H, \dots, T)$. 

If  $F = N$, it corresponds to a unique microstate $(H, H, H, H, \dots, H)$
    
But if   $F = \frac{N}{2}$ then there are approximately  $\frac{2^N}{\sqrt{N}}$  microstates giving rise to that value of $F$.

If one considers the density function, it is easy to see that if the box $\Lambda$ is divided in two equal parts, the volume in phase space where the particles are uniformly distributed in $\Lambda$ will be of the order of $2^N$ times larger than the one where  the particles are concentrated in one of those parts.

\begin{figure}[!ht]
\centering
\includegraphics[width=.7\textwidth]{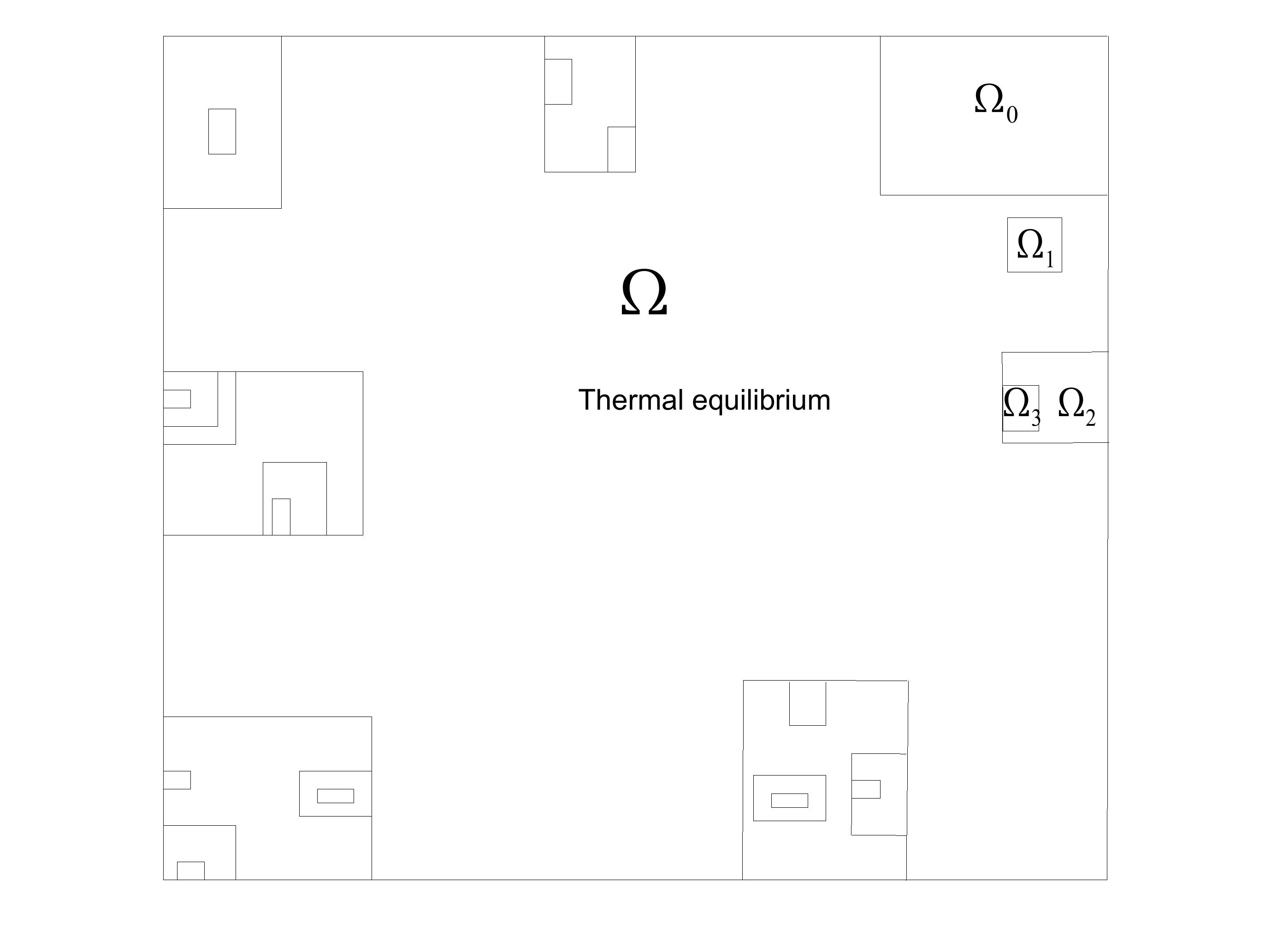}
\caption[]{A partition of the phase space $\Omega$ (represented by the entire square) into regions $\Omega_0$, $\Omega_1$, $\Omega_2, \dots$ corresponding to microstates that are macroscopically indistinguishable from one another,
i.e. that give rise to the same value of $F$. The region labelled ``thermal equilibrium" corresponds to the value of $F$ corresponding to the overwhelming majority of microstates.}\label{fig1}
\end{figure}

\begin{figure}[!ht]
\centering
\includegraphics[width=.7\textwidth]{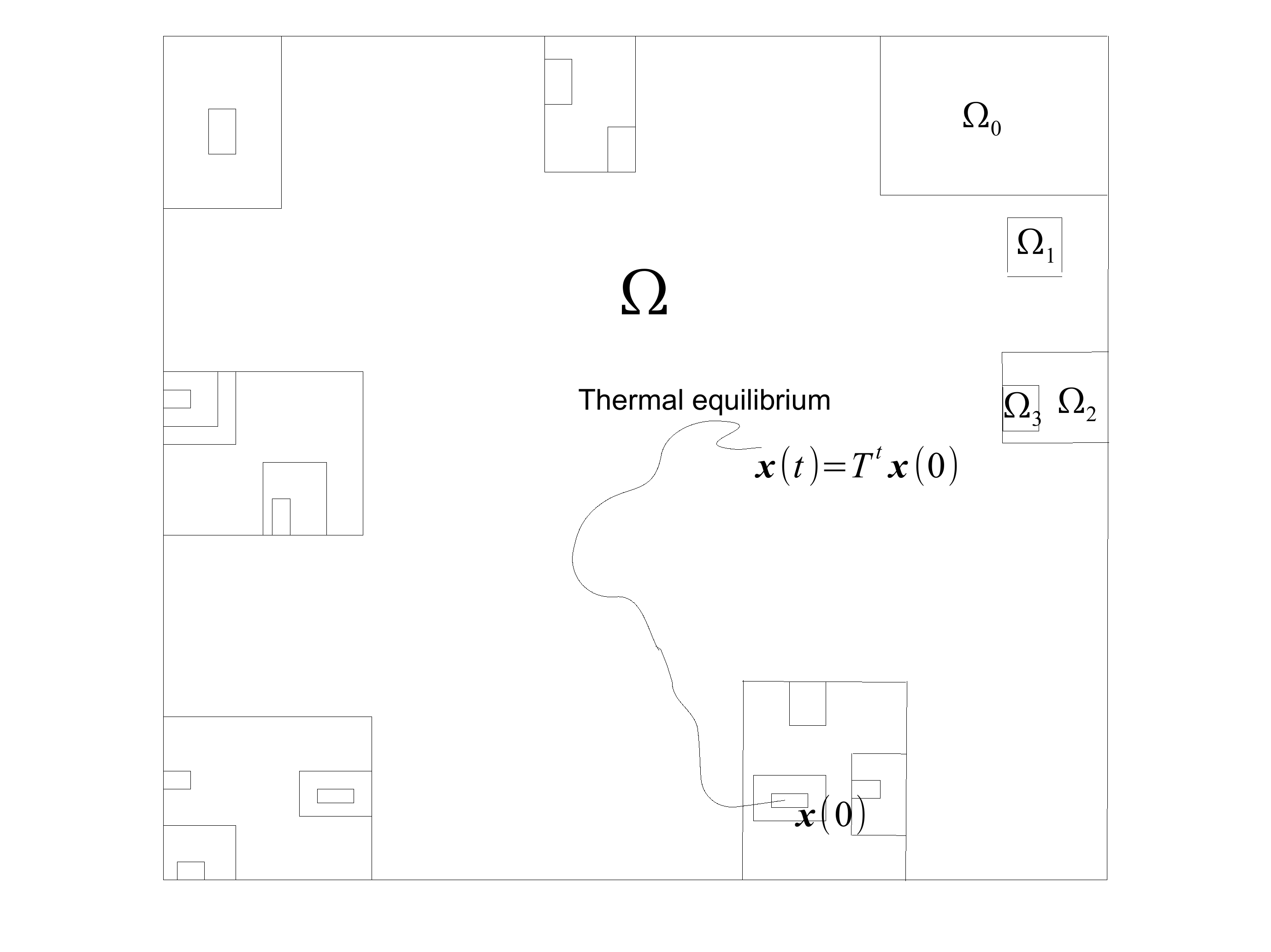}
\caption[]{The curve ${\bf x}(t)=T^t {\bf x}(0)$ describes a possible evolution of a microstate, which tends to enter regions of larger volume until it enters the region of thermal equilibrium.}\label{fig2}
\end{figure}

\begin{figure}[!ht]
\centering
\includegraphics[width=.7\textwidth]{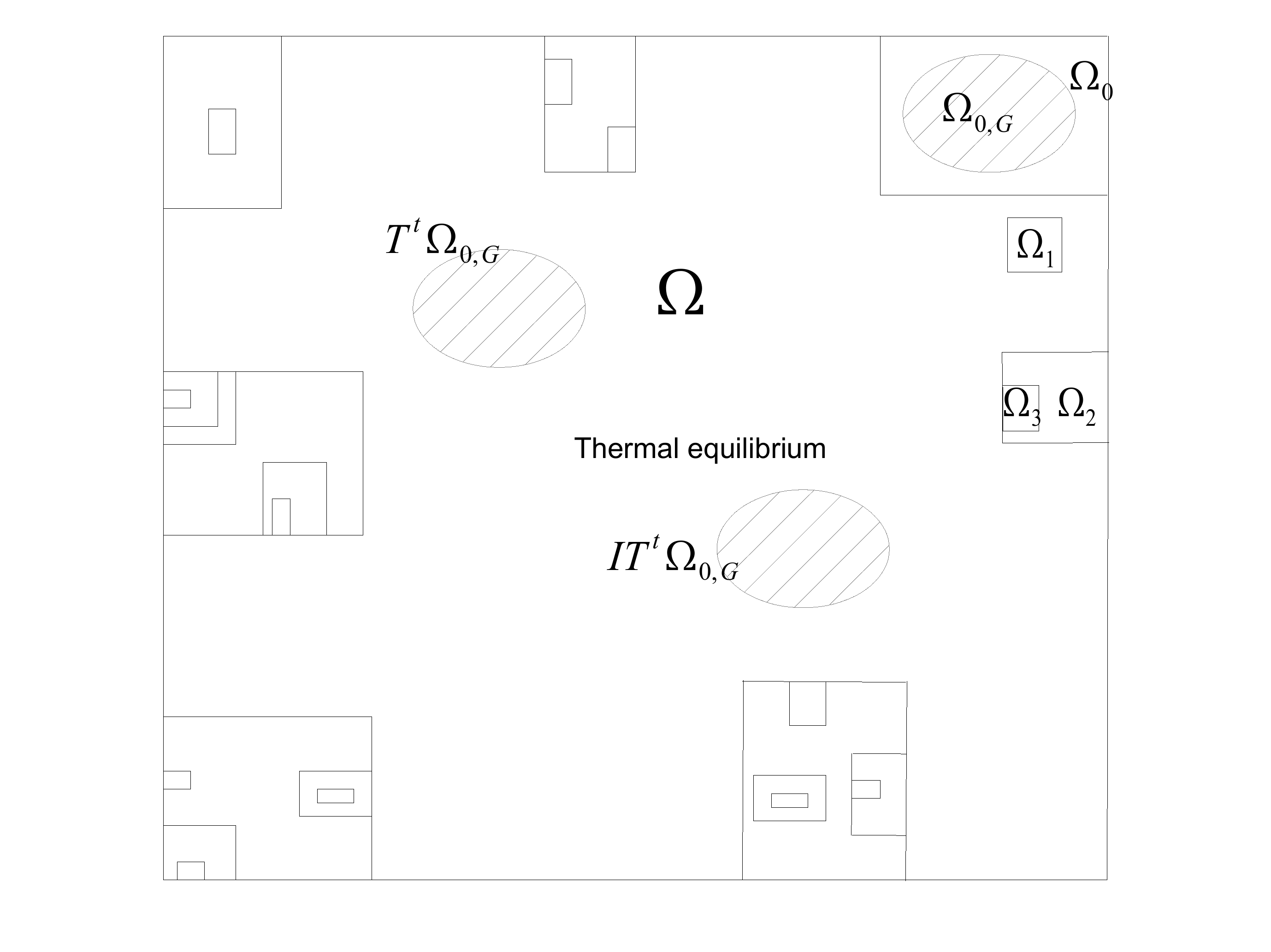}
\caption[]{$\Omega_0$ are the
configurations
in one-half of the box at time zero;
 $\Omega_{0, G} $  are the good  configurations in $\Omega_0$  
whose evolution lead to a uniform density at time $t$: $T^t (\Omega_{0, G}) \subset \Omega_t$;
$I(T^t (\Omega_{0, G}))$ are the configurations of $T^t (\Omega_{0, G})$ with velocities reversed whose 
evolution after time $t$ belongs to $\Omega_0$: $T^t
(I(T^t (\Omega_{0, G})) \subset \Omega_{0}$. 
}\label{fig3}
\end{figure}

In figure \ref{fig1}, one has a schematic illustration of this fact\footnote{Figures \ref{fig1}, \ref{fig2}, \ref{fig3} are inspired by similar pictures in chapter 7 of \cite{Penr}.}: each region in  $\Omega$ corresponds to the set of microstates giving 
rise to the same value of $F$ and we denote these regions by $\Omega_0$, $\Omega_1$, $\Omega_2,\dots$.
Let us stress that the figures \ref{fig1}, \ref{fig2}, \ref{fig3} are highly ``abstract" since the phase space $\Omega$
 represented there by a two-dimensional square is in reality a subset of a space of dimension of order $10^{23}$.

Moreover, $F$ usually takes a continuum of values and in figure \ref{fig1} we do as if those values were discrete. But that is only in order to simplify our illustration.

In the example of coin tossing, the region labelled  $\mbox{equilibrium}$ in figure \ref{fig1} corresponds to having approximately as many heads and tails, i.e. the set of coin tossings defined by (\ref{G}). The region $\Omega_0$ may correspond to having approximately one third heads and two third tails, $\Omega_1$ may correspond to having approximately one quarter heads and three quarters tails, etc.

In the example of the gas in the box,  the  region labelled ``thermal equilibrium" in figure \ref{fig1} corresponds to an approximate uniform  distribution of the particles in the box and an approximate Maxwellian distribution of their velocities.\footnote{We use the word ``approximate" here, because, as for coin tossing, for a finite number of particles, the correspondence with the predicted statistical distribution is always approximate.} 

The region $\Omega_0$ in figure \ref{fig1} might correspond to all the particles being in one half of the box, another region might correspond to all the particles being in the other half, yet another region, say $\Omega_1$,  might correspond to all the particles being in an even smaller part of the box etc.

Of course, nothing is drawn to scale here: if the size of the region in the phase space where   all the particles are concentrated in one part of the box is $2^{-N}$ smaller than the one where the particles are uniformly distributed in $\Lambda$, and $N$ is of the order of the Avogadro's number, $N\sim 10^{23}$, that region  where   all the particles are concentrated in one part of the box could not be seen at all if things were drawn to scale.

 The thermal equilibrium region is almost equal to the entire phase space $\Omega$ and all the non-equilibrium regions put together (all the particles concentrated in one part of the box, or the distribution of the velocities being different  from the Maxwellian one) occupy only a tiny fraction of $\Omega$.

To understand how $F_t$ can evolve irreversibly even though its evolution is induced by a reversible microscopic evolution, consider figure \ref{fig2}, which illustrates what  one expects to happen: the microstate ${\bf x}(t)$ evolves towards larger and larger regions of phase space and eventually ends up in  the  ``thermal equilibrium" region. Therefore, the induced evolution of $F_t= F ({\bf x}(t))$ should tend towards equilibrium. However, at the level of generality considered here, our expectation is simply based on the fact that some regions are (much) bigger than others, and so it would be natural for the microstate ${\bf x}(t)$ to evolves towards those bigger regions if nothing prevents it from doing so. 

There are several caveats here: one is that this scenario is what one expects or hopes for. We will give in the next section an example, a rather artificial one, where this scenario can be demonstrated in detail, which shows that this scenario is certainly possible and even plausible, but it is certainly not demonstrated in any degree of generality in physically natural situations. 

The more important caveat is that, even if this scenario is true, the desired evolution is definitely not true for all microstates  ${\bf x}(t)=T^t {\bf x}(0)$ giving rise to a given value $F_t= F ({\bf x}(t))$. That follows  from the reversibility argument given in subsection \ref{sec6.1}: for every 
microstate  ${\bf x}(t)$ that induces  the irreversible  evolution of $F_t= F ({\bf x}(t))$ there exists another microstate
$I({\bf x}(t))$ so that  $F_t= F(I({\bf x}(t)))$  at time $t$, but $I({\bf x}(t))$
induces  a different evolution than the irreversible one for times later than $t$.

Let us illustrate this explanation of irreversibility
in a concrete physical example. Consider the gas
introduced in subsection \ref{sec6.1} that is
initially compressed by a piston in one-half of the box $\Lambda$, and that
expands  into the whole box. Let $F$ be the density
of the gas. Initially, it is equal to $1$ (say) in one half of the box and to $0$ in the
other half. After some time $t$,
 it
is (approximately) equal to
$\frac{1}{2}$ everywhere. The explanation of the
irreversible evolution of $F$ is
 that the overwhelming majority of the microscopic configurations
corresponding to the gas in one-half of the box, will evolve deterministically
so as to induce the observed evolution of $F$. There may of course be some
exceptional configurations, for which all the particles stay in the left half.
All one is saying is that those configurations are extraordinarily rare,
and that we do not expect to see even one of them appearing when
we repeat the experiment many times.
So,  the microscopic configurations that lead to the
expected macroscopic behavior will  be {\it typical} in the sense of (\ref{typ}). 

Let us define the {\it good configurations}  (up to a certain time $T$) as being those configurations that induce the macroscopic law up to time $T$, and the {\it bad configurations} (up to a certain time $T$) the other configurations; we will use indices $G$ and $B$ for the corresponding sets of configurations.\footnote{We introduce the time upper bound  $T$ here, because, if we wait long enough, all configurations will be bad, since the Poincar\'e's recurrence theorem (see e.g. \cite{Ar}) implies that they will all come back arbitrarily close to their initial conditions. In the example of the gas in the box, this means that all the particles  will come back simultaneously to the half-box in which they were initially, which is contrary to the behavior predicted by the macroscopic laws. But the time needed for a  Poincar\'e's recurrence to occur in a large system is typically much larger than the age of the Universe, so that, from a practical point of view, we may take $T=\infty$.} 

Now, take all the good microscopic
configurations in one-half of the box, and let them evolve up to a time $t <<T$, when 
the density is approximately uniform.
Now, reverse all the velocities. We get a set of configurations that still
determines a density approximately $\frac{1}{2}$ in the box, so the value of the density function $F$, defined by (\ref{macro1}), is unchanged. However, those configurations  are not good any more.
Indeed, from now on, if the system remains isolated, the density just remains
uniform according to the macroscopic laws. But for the
 configurations just described, the gas will move back
to the part of the box that they started from,  see (\ref{rev}), leading to a gross violation of the macroscopic law. What is
the solution? Simply that those ``reversed-velocities" configurations form
a very tiny subset of all the microscopic configurations giving rise to a
uniform density. And, of course, the original set of configurations, those
coming from the initial half of the box, also form such a small subset.
Most configurations corresponding to a uniform density
do not go to one-half of the box, either in the future or in the past.
 
To express this  idea in formulas, let  $ \Omega_t $ be the set of
the configurations giving to the function $F$ its value $F_t$ at time $t$.
In other words,  $\Omega_t $ is
 the pre-image of $F_t$ under the map $F$.
Let
 $\Omega_{t, G}$ be the set of {\it good configurations}, at
time $t$, namely those that  lead to a
behavior of $F$  following the macroscopic
laws,  up to some time $T>>t$.

One expects that, in general, $\Omega_{t, G}$ is a  very large
subset of $ \Omega_t$ (meaning that $\frac{|\Omega_{t, B}|}{|\Omega_t|}=\frac{|\Omega_t \setminus \Omega_{t, G}|}{|\Omega_t|}<<1$), but is not identical to $\Omega_t$ (see (\ref{typ2}) below for a more precise statement).

In our example of the gas initially compressed in one-half of the box $\Lambda$, the set $\Omega_0$ consists of all the
configurations
in one-half of the box at time zero, and
 $\Omega_{0, G} $ is the subset consisting of those configurations
whose evolution lead to a uniform density at time $t$, which means that $T^t (\Omega_{0, G}) \subset \Omega_t$.

Microscopic reversibility says that $T^t
(I(T^t (\Omega_{0, G}))=I(\Omega_{0, G}) \subset I(\Omega_{0})=\Omega_{0}$ 
(this is just (\ref{rev}) 
applied to the set $\Omega_{0, G}$). The last equality holds because the set of configurations
in one-half of the box  is invariant under the change of the sign of the velocities.

A  paradox
would occur if $T^t(I(\Omega_t)) \subset I(\Omega_{0})=\Omega_0$. Indeed, this would mean that, if 
one reverses the velocities of all the configurations at time $t$ corresponding to a uniform density, and  let
them evolve for a time $t$, one would get a set of configurations  in one-half of the box (with velocities reversed). Since the operation $I$ preserves the Lebesgue measure $|I(\Omega_t)|= |\Omega_t|$, this would imply that there are as many configurations corresponding to a uniform density (configurations   in $\Omega_t$)
 as
there are configurations that will evolve back to one-half of the box in time $t$ (those in $I(T^t (\Omega_{0, G})$). Or, in other words, one would have  $I(\Omega_t) \subset\Omega_{t, B}=\Omega_t \setminus \Omega_{t, G}$, which combined with $|I(\Omega_t)|= |\Omega_t|$ makes the bound
$\frac{| \Omega_{t, B}|}{|\Omega_t|}=\frac{|\Omega_t \setminus \Omega_{t, G}|}{|\Omega_t|}<<1$  impossible.

But $\Omega_t$ is {\it not at all equal}, in general,
 to $T^t
(\Omega_{0, G})$, so that $T^t
(I(T^t (\Omega_{0, G}))=I(\Omega_{0, G}) \subset I(\Omega_{0})=\Omega_{0}$ {\it does not imply}
 $T^t(I(\Omega_t)) \subset I(\Omega_{0})=\Omega_0$. 
In our  example, $T^t
(\Omega_{0, G})$ is a tiny subset of $ \Omega_t$, because most configurations
in $\Omega_t$ were not in half of
the box at time zero. 

This is illustrated in figure \ref{fig3}: $\Omega_0$ is the set of configurations with all the particles in one-half of the box, and
$\Omega_{0, G}$ the subset of those that evolve towards a uniform distribution after some time $t$. Thus $T^t (\Omega_{0, G})$ is a  subset of $ \Omega_t$, which is the set of thermal equilibrium states. Reversing the velocities of every configurations in $T^t (\Omega_{0, G})$ yields the set $I(T^t (\Omega_{0, G}))$, which is also a subset of $ \Omega_t$, but a ``bad" subset, namely one that does not stay in equilibrium but moves back to the half box where the particles were to start with. So, $F$ applied to those configurations will not evolve according to the usual macroscopic laws (which implies that the density stays uniform in the half box), which is what we mean by bad configurations.

Of course it should be emphasized once more that the subsets in figure \ref{fig3} are not drawn to scale: the sets $\Omega_0$, $T^t (\Omega_{0, G})$ and  $I(T^t (\Omega_{0, G}))$ are minuscule compared to the set of equilibrium configurations $ \Omega_t$.

In fact, one knows from Liouville's theorem\footnote{Which says that the Lebesgue measure of a set is invariant under the Hamiltonian flow $T^t$, see e.g. \cite{Ar}.} that the size of $\Omega_{0, G}$  and  $T^t (\Omega_{0, G})$  are equal: $|\Omega_{0, G}|=| T^t (\Omega_{0, G})|$. Since the operation $I$ also preserves the size of a set, we have:  $|\Omega_{0, G}|=| T^t (\Omega_{0, G})|=|I(T^t (\Omega_{0, G}))|$, which is illustrated in figure \ref{fig3}. 

Since $\Omega_{0, G}\subset \Omega_0$, we have $|\Omega_{0, G}|\leq | \Omega_0| $ and we already observed that the ratio  $\frac{| \Omega_0|}{ | \Omega_t|  } \sim 2^{-N}$ (since each of the $N$ particles can be in either half of the box in $\Omega_t$, but only in one-half of the box in $ \Omega_0$).  Thus, 
\be
\frac{| I(T^t(\Omega_{0, G}))|}{ | \Omega_t|} = \frac{| T^t(\Omega_{0, G})|}{ | \Omega_t|} = \frac{|  \Omega_{0, G}|}{ | \Omega_t|}  \leq\frac{|  \Omega_0|}{ | \Omega_t|} \sim 2^{-N},
\ee
 which is astronomically small for $N\sim 10^{23}$.

 What one would like to show is that the good configurations are {\it typical} in the sense of (\ref{typ}). More precisely, one want to show that, for all times $T$ not too large, and all $t<<T$,
 \be
\frac{| \Omega_{t, G}|}{ | \Omega_t|}\to 1
\label{typ2}
\ee
as $N\to \infty$.

\subsection{Irreversibility and probabilistic explanations}\label{sec6.3}

The above explanation of the irreversible behavior of $F_t$ is again probabilistic: the vast majority of microstates
${\bf x}(0)$ corresponding to
the macrostate $F_0$  induce, through their deterministic evolution  ${\bf x}(0) \to T^t({\bf x}(0))= {\bf x}(t)$,  the expected time evolution of $F_0 \to F_t$. 

What else could one ask for? One could wish to show that the expected time evolution of $F_0 \to F_t$ is induced by {\it all } microstates
${\bf x}(0)$ corresponding to
the macrostate $F_0$. But we showed by explicit counterexamples that,  in general,  this is not possible.

So, our explanation is the best one can hope for. But is it satisfactory? It is, provided that one accepts the notion of probabilistic explanation 
given in section \ref{sec5}. And if one does not accept it, it is not clear what notion of explanation one has in mind and 
how one could justify it.
Nevertheless,  by speaking of a ``vast majority of microstates", we did not say what we meant  by the ``vast majority". Since
there are uncountably many such states one needs a measure on $\Omega$ in order to make sense of that notion.

The measure on subsets of  $\Omega$  that we used implicitly here is the size of the set or its Lebesgue measure.\footnote{If $\Omega$  is a set of measure zero in $\R^{6N}$, for example a constant energy surface, one has to consider the restriction of the  Lebesgue measure to that surface instead of the  Lebesgue measure on  $\R^{6N}$. This measure is called the Liouville measure.}
But one could still ask: why  use that measure and not some other measure? Obviously if one could not argue that this measure is in some sense ``natural", our whole notion of explanation would collapse. Indeed, it is easy to invent measures that will give a much greater weight to, for example, the set $\Omega_0 \setminus \Omega_{0, G}$ than to $ \Omega_{0, G}$. Then, if we define the probability of a set according to such a measure,  the usual induced evolution $F_0 \to F_t$ becomes improbable and some other evolution becomes probable.

So something has to be said in favor of the naturalness of the  Lebesgue measure. But that is easy enough on Bayesian grounds or on the basis of the indifference principle: the Lebesgue measure is the most symmetric one, being invariant under translations and rotations and varying naturally under scalings. So, from that point of view, there is no alternative to taking the  Lebesgue measure as our natural measure and, if the set of initial states inducing the expected evolution of $F_t$ has a  high probability relative to that measure (i.e. are typical relative to that measure), then we will consider that the evolution of  $F_t$ has been explained by the microscopic laws. 

Some people want to justify the naturalness of the Lebesgue measure by invoking the fact that it is time-invariant under the Hamiltonian flow (Liouville's theorem). While this is true, it is not necessarily the best argument in favor of the naturalness of the Lebesgue measure. Indeed there exists non-Hamiltonian systems, for which the Lebesgue measure is {\it not } invariant and that 
do possess an invariant measure $\nu$ whose support is a set $\cal A$ of zero Lebesgue measure. Moreover, one can, in certain cases, prove that, for almost every initial condition ${\bf x}_0$ with respect to the Lebesgue measure on a set of non-zero Lebesgue measure containing $\cal A$, the time evolution ${\bf x}_0 \to {\bf x}_t$  drives the trajectory towards $\cal A$ and the statistics of 
the time spent  by  the trajectory close to subsets of $\cal A$ is proportional to the $\nu$-measure of those subsets.\footnote{We think here of certain ``chaotic" dynamical systems for which $\nu$ is a Sinai-Ruelle-Bowen measure and the set $\cal A$  is a ``strange attractor", see e.g. \cite{Bow, BR, ER, Sin, Ru1, Ru2}.}

In the following section we will illustrate  the previous ideas in a concrete situation.

\section{The Kac ring model}\label{sec7}

\subsection{The model}\label{sec7.1}

Let us consider a simple model, due to Mark Kac
(\cite{Ka} p.99, see also
Thompson (\cite{Tho} p.23) and \cite{GO}), which nicely
illustrates Boltzmann's solution to the problem of
irreversibility, and shows how
to avoid various misunderstandings and paradoxes.

We will use a slightly modified version of the model
 and state the relevant results,
 referring to \cite{Ka}
for the proofs.

\begin{figure}[!ht]
\centering
\includegraphics[keepaspectratio,height=8cm]{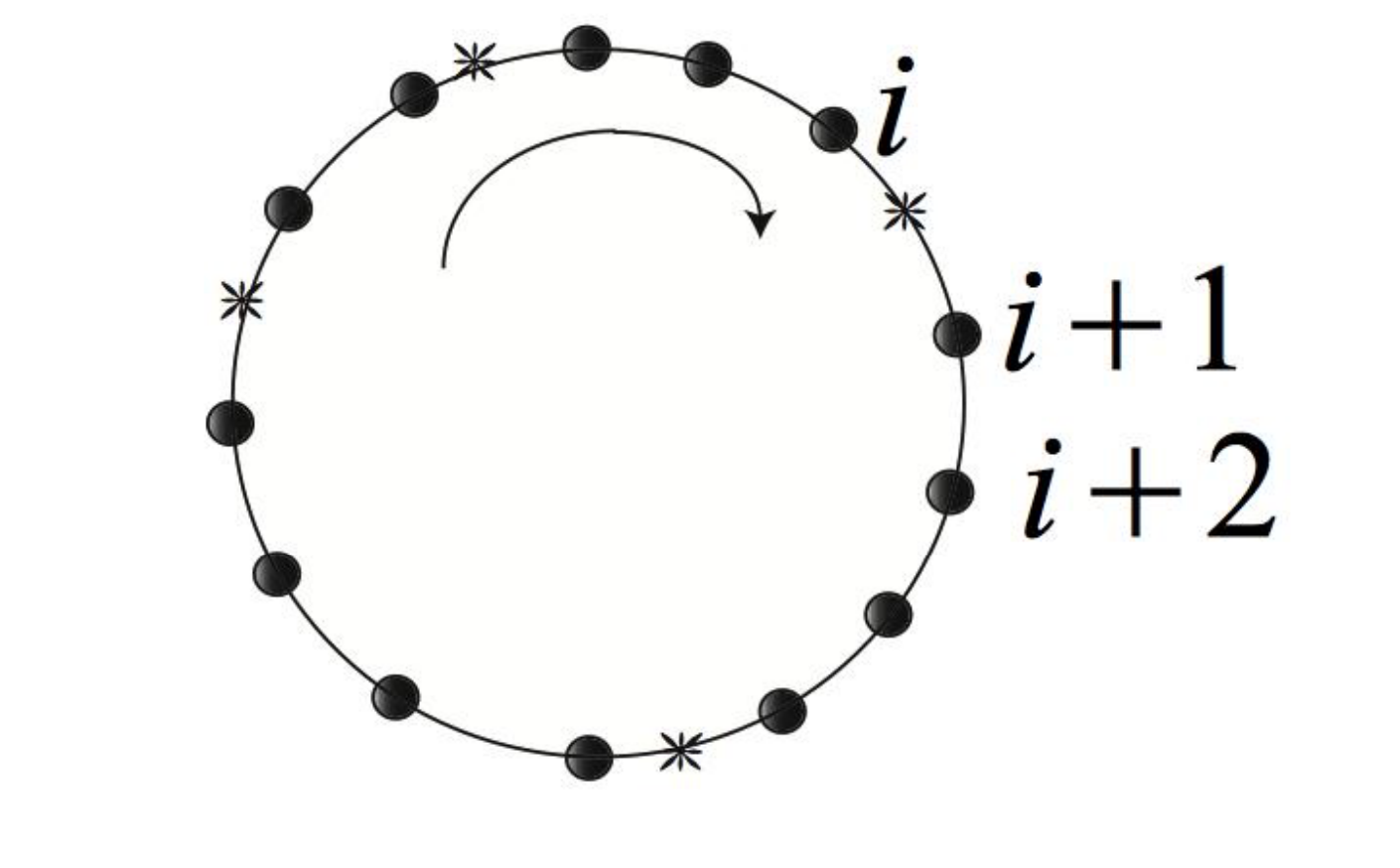}
\caption[]{At each site $i$ there is a particle that has either a plus or minus sign. During an elementary time interval each particle
 moves clockwise
to the nearest site.
If the particle crosses an interval marked with a cross (as the one between the sites $i$ and $i+1$), it changes sign
but if it crosses an interval without a cross (as the one between the sites $i+1$ and $i+2$) it
does not change its sign.}\label{fig4}
\end{figure}

 One considers $N$ equidistant points on a circle; $M$  of the
 intervals between
the points are marked and form a set called $S$. The complementary set
 (of $N-M$
intervals) will be called
$\bar S$.
We will define 
\be
\alpha= \frac{M}{N}.
\label{alpha}
\ee
It will be convenient later to assume that
\be
\alpha <\frac{1}{2}.
\label{7.4}
\ee

Each of the $N$ points there is a particle that can have 
either a plus sign  or
minus sign (in the original Kac model, one speaks of white and black balls). During an elementary time interval each particle
 moves clockwise
to the nearest site, obeying the following rule:
if the particle crosses an interval in $S$, it changes sign
upon completing the move
but if it crosses an interval in
$\bar S$, it performs the move without changing sign.

Suppose that we start with all  particles having a  plus sign; the question
is what happens after a
large number of moves.
After  (\ref{7.3}) we shall also consider other initial conditions.

To formalize the model, introduce for each $i=1,\dots, N$, the variable\footnote{See for example figure \ref{fig4} where $\epsilon_{i}=-1$ and $\epsilon_{i+1}=+1$.}
\bea
\epsilon_i = \left\{ \begin{array}{c} +1 \; \mbox{if the interval in front of} \; i
\in
\bar S
\\ -1 \; \mbox{if the interval in front of} \; i \in S \end{array} \right.
\eea
and we let $\eta_i (t) =\pm 1$ be the sign of the particle at site $i$ and time $t$.

Then, we get the ``equations of motion":
\be
\eta_{i+1} (t+1) = \eta_{i} (t) \epsilon_{i}.
\label{eq}
\ee

Let us first explain the analogy with mechanical laws. The particles are described by
their positions and their (discrete) ``velocity", namely their sign. One of the
simplifying features of the model is that the ``velocity" does not affect the
motion. The only reason one calls it a
 ``velocity" is that it changes when the particle collides with a fixed
``scatterer", i.e. an interval in $S$. Scattering with fixed objects
tends to be
easier to analyse
than collisions between particles. The ``equations of motion" (\ref{eq}) are given by
the clockwise motion, plus the changing of signs.
These equations are obviously deterministic and reversible: if, after a time $t$, we
change the orientation of the motion from clockwise to
counterclockwise, we return
after $t$ steps to the original state.\footnote{There is a small
abuse here, because it seems that we  change the laws of motion by changing
the orientation (from clockwise to
counterclockwise). But one can attach another discrete ``velocity" parameter
to the particles, having the same value for all of them, and
indicating the orientation, clockwise or counterclockwise, of
their motion. Then, the motion is truly
reversible, and we have simply to assume that the analogue here of the operation $I$ of  (\ref{I})
changes also that extra
velocity parameter.} Moreover, the motion is strictly periodic:
after $2N$ steps each interval has been crossed twice by each particle, hence they all
come back to their original sign.\footnote{This is analogous to the Poincar\'e cycles in mechanics,
except that, here, the length of the cycle is the same for all
configurations (there is no reason for this feature to hold in general mechanical
systems).} 

\begin{figure}[!ht]
\centering
\includegraphics[keepaspectratio,height=8cm]{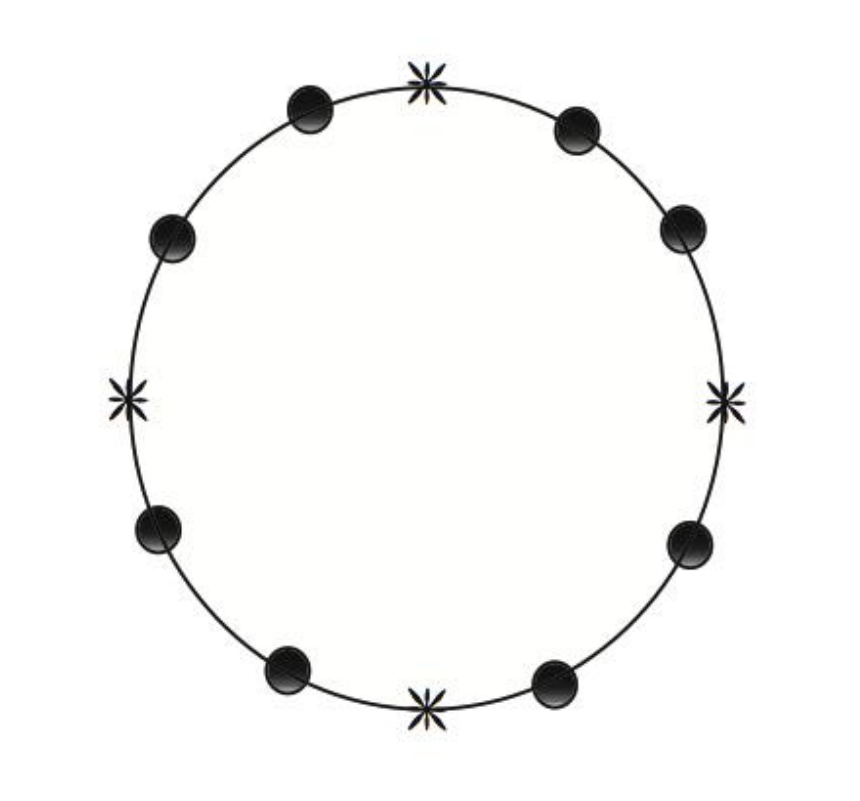}
\caption[]{An example of distribution of crosses where every configuration is periodic of period 4.}\label{fig5}
\end{figure}

It is also easy to find special configurations which obviously do
not tend to equilibrium: start with all  particles being ``plus" and
let every other interval
belong to $S$ (with $M=\frac{N}{2}$). Then, after two steps, all particles are minus,
after four steps they are all plus again, etc... The motion is periodic with
period 4, see figure \ref{fig5} for a simple example.

Turning to the solution, one can start by analyzing the
approach to equilibrium in this model \`a la
Boltzmann.

\subsection{Analogue of  Boltzmann's solution.}\label{sec7.2}

Let $N_+(t)$, $(N_-(t))$ denote
the total number of ``plus"  particles or ``minus"  particles at time $t$ (i.e., after $t$ moves; $t$
being an integer). $N_+(t)$ and  $(N_-(t))$ are the macroscopic variables in this problem (since $N_+(t)+(N_-(t))=N$
there is only one independent macroscopic variable).

Let $N_+(S;t)$, $N_-(S;t))$ be the number of  ``plus" particles or  of  ``minus" particles
which are going to cross an interval in $S$ at time $t$.

We have the immediate conservation relations:
\ba
N_+(t+1) &=& N_+(t) - N_+(S;t) + N_- (S;t) \nonumber \\
N_- (t+1) &=& N_-(t) - N_- (S;t) + N_+ (S;t) \label{7.1}
\ea

If we want to solve (\ref{7.1}), we have to make some
assumption about $N_+(S;t)$, $N_- (S;t)$. Otherwise, one has to write down equations for
$N_+(S;t)$, $N_- (S;t)$ that will involve new variables and lead to a potentially
infinite regress.

So,  following Boltzmann, we introduce the assumption (``Stosszahlansatz" or
``hypothesis of molecular chaos"\footnote{The word ``chaos" here has nothing
to do with ``chaos theory", in the sense of dynamical systems (see footnote 16), and, of course,
Boltzmann's hypothesis is much older than
that theory.}):
\ba N_+(S;t) &=& \alpha N_+ (t) \nonumber \\
N_- (S;t) &=&  \alpha N_- (t), \label{7.2}
\ea
with $\alpha$ defined in (\ref{alpha}).

The intuitive justification for this assumption is that each particle is
``uncorrelated" with the event ``the interval ahead of the particle   belongs to
$S$", so we write $N_+ (S;t)$ as equal to $N_+(t)$, the total number of  ``plus"  particles, times the density $\alpha$ of intervals in $S$. This assumption looks
completely reasonable. However, upon reflection, it may lead to some puzzlement: what does ``uncorrelated"
exactly mean? Why do we introduce a statistical assumption in a mechanical model?
Fortunately here, these questions can be answered precisely and we shall answer
them later by solving the model exactly. But let us return to the Boltzmannian
story.

One obtains from (\ref{7.2}):
$$
N_+(t+1) - N_- (t+1) = (1-2 \alpha)(N_+(t)-N_-(t))
$$
Thus
\ba
N^{-1} [N_+ t) - N_- (t)] &=&  (1-2 \alpha)^t N^{-1}[N_+(0)-N_-(0)],\nonumber\\
&=& (1-2 \alpha)^t. \label{7.3}
\ea
since $N_+(0)=N$, $N_-(0)=0$. 

Using  (\ref{7.4}) and (\ref{7.3}), we obtain a {\em monotonic} approach to
equal number of particles ``plus" and   particles ``minus", i.e. to equilibrium. Note that we get a monotonic
approach for {\em all} initial conditions ($N_+(0)-N_-(0)$) of the particles.

We can see here in what sense Boltzmann's solution is an approximation. The
assumption (\ref{7.2}) cannot hold for all times and for all
configurations, because it
would contradict the reversibility and the periodicity of the motion. However,
we will show now that the fact that it is an approximation does not invalidate
Boltzmann's ideas about irreversibility.

\subsection{Microscopic analysis of the model.}\label{sec7.3}

Let us reexamine the model at the microscopic level, first
mechanically and then
statistically. The  solution of the equations of motion (\ref{eq}) is:
\be
\eta_i (t) = \eta_{i-t} (0) \epsilon_{i-1} \epsilon_{i-2} \cdots \epsilon_{i-t}
\label{7.6}
\ee
(where the subtractions in the indices are done modulo $N$). So we have an explicit solution of
the equations of motion at the microscopic level.

We can express the macroscopic variables in terms of that solution:
\be
N_+ (t) - N_-(t) = \sum_{i=1}^n \eta_i (t) = \sum^n_{i=1} \eta_{i-t} (0)
\epsilon_{i-1} \epsilon_{i-2} \cdots \epsilon_{i-t}
\label{7.7}
\ee
and we want to compute $N^{-1} (N_+ (t) - N_- (t))$ for large $N$, for various
choices of initial conditions $\{ \eta_i (0)\}_{i=1}^N$ and various sets $S$
(determining the $\epsilon_i$'s). It is  here that ``statistical" assumptions
enter. Namely, we fix an arbitrary initial condition $\{ \eta_i (0)\}_{i=1}^N$
 and consider all possible
sets $S$ with $M=\alpha N$ fixed (one can of course think of the choice of $S$ as
being part of the choice of initial conditions). Then, for each set $S$, one
computes the ``curve" $N^{-1} (N_+ (t) - N_- (t))$ as a function of time. 

The
result of the computation, done in \cite{Ka}, is that, for any given $t$ and for
$N$ large, the overwhelming majority of these curves will
approach $(1-2\alpha)^t$, i.e. what is predicted by (\ref{7.3}).
(to fix ideas, Kac suggests to think of $N$ as being of the order $10^{23}$ and
$t$ of order
$10^6$). The fraction of all curves that will deviate significantly from
$(1-2\alpha)^t$, for fixed  $t$, goes to zero as $N^{-\frac{1}{2}}$, when $N\to
\infty$.

Of course when we say ``compute", one  should rather say that one makes an estimate of
the fraction of curves deviating from $(1-2\alpha)^t$ at a fixed $t$.
This estimate is similar to the law of large numbers since (\ref{7.7}) is  of the form
of a sum of (almost independent) random  variables.

Let us express what happens in this model  in terms of the sets depicted in figure \ref{fig3}.
The ``phase space" $\Omega$ consists of all configurations of signs and scatterers (with $2M<N$), $\{\epsilon_i, \eta_i\}_{i=1}^N$.

The ``thermal equilibrium" set in figure \ref{fig3}  corresponds to the set of 
configurations of particles such that  $N^{-1} (N_+ (t) - N_- (t))$ is approximately equal to $0$ and to the set of configurations of scatterers  such that 
 $N^{-1} (N_+ (t) - N_- (t))$ remains close to $0$ in the future.
 
It is easy to compute  the number of (microscopic)
configurations whose number of   ``plus" particles is $N_+(t)$. It is given by:
\be
\left( \begin{array}{c} N \\
N_+(t) \end{array} \right) =\frac{N!}{N_+(t)! (N-N_+(t))!}
\label{number}
\ee
and this number reaches its maximum value for $N_+ = \frac{N}{2} = N_-$.

We can, as in figure \ref{fig1},  introduce a partition
of the phase space according to the different values of $N_+$, $N_-$.
And what (\ref{number}) shows is that different elements of that partition
have very different number of elements, the vast majority
corresponding to  ``equilibrium", i.e. to those
 near $N_+ = \frac{N}{2} = N_-$.

If one illustrates this model through  figure \ref{fig3}, the set $\Omega_0$ consists of all configurations of scatterers and of all particles being ``plus": $\eta_i=+1$, $\forall i= 1, \dots, N$.
The subset $\Omega_{0, G} \subset \Omega_0$ of good configurations consists of those configurations of scatterers such that
$N^{-1} (N_+ (t) - N_- (t))$ tends to $0$ and of all the particles having a plus sign.

 Then $T^t(\Omega_{0, G})\subset \Omega_t$ is a set of configurations with $N^{-1} (N_+ (t) - N_- (t))$ approximately equal to $0$ but a set of scatterers that is special in the following sense: the  configurations in $I(T^t(\Omega_{0, G}))$, where $I$  changes the orientation of the motion from clockwise to counterclockwise,
will evolve in a time $t$ to a configuration with all particles being plus.

So, although, as far as the signs of the particles are concerned, there is nothing special about the configurations in $T^t(\Omega_{0, G})$ ($N^{-1} (N_+ (t) - N_- (t))$  is close to $0$), there is a subtle correlation between the configurations of the particles and the scatterers  in $T^t(\Omega_{0, G})$, as shown by what happens if one applies the orientation-reversal operation $I$ to those configurations. This is simply a ``memory effect" due to the fact that the configurations in $T^t(\Omega_{0, G})$ were initially in 
$\Omega_0\supset \Omega_{0, G}$ (this is similar to the memory effect of the particles of the gas in section \ref{sec6} that were initially in one half of the box).

But the configurations in $T^t(\Omega_{0, G})$ form a very small subset of $\Omega_t$; indeed, $|T^t(\Omega_{0, G})|=|\Omega_{0, G}|\leq |\Omega_0| $, and  $\frac{|\Omega_0|}{|\Omega|}= 2^{-N}$ (because there are two possible signs in $\Omega$ for each of the $N$ sites, but only one sign, plus, in $\Omega_0$).  By the law of large numbers one can show that $|\Omega_t| \sim |\Omega|$ for $N$ large, so that, $\frac{|\Omega_0|}{|\Omega_t|}\sim 2^{-N}$, again for $N$ large and thus $\frac{|\Omega_0|}{|\Omega_t|}$ is extremely small in that limit.

Note that here, we define typical behavior by counting the number of configurations, see (\ref{number}), which is the same as putting a probability equal to $\frac{1}{2}$
to each particle sign (note also that, in the estimates made on (\ref{7.7}), we had implicitly  put a probability equal to $\frac{1}{2}$ to the presence or not of a  scatterer on each interval). This uniform probability is again the one following from the indifference principle. 

I do not want to overemphasize the interest of the Kac model. It
has many simplifying features (for example, there is no conservation
of momentum; the scatterers here are ``fixed").
However, it has $all$ the properties that have been invoked to show that
mechanical systems cannot behave irreversibly, and therefore it is a
perfect counterexample that allows us to refute all those arguments (and to
understand exactly what is wrong with them): it is isolated (the particles plus
the scatterers), deterministic, reversible and periodic.\footnote{Periodicity is a stronger property than the existence
 of Poincar\'e cycles and implies that the system is not
ergodic. For a discussion of why ergodicity is neither necessary to sufficient  in order to justify approach to equilibrium, see \cite[section 4.2]{Bri}.}

This result, obtained in the Kac model, is exactly what one would
like to show for general mechanical systems,
in order to establish irreversibility. It is obvious why this is very hard. In
general, one does not have an explicit solution (for an $N$-body system!) such as
(\ref{eq}, \ref{7.6}), in terms of which the macroscopic variables can be expressed, as in (\ref{7.7}).

If
we prepare a Kac model many times and if the only variables that we can control
are $N$ and $M$, then we expect to see the irreversible behavior obtained
above, simply because this is what happens {\it deterministically} for the vast
majority of microscopic initial conditions corresponding to the
macroscopic variables that we are able to
control.

\subsection{Conclusions}\label{sec7.4}

In this paper, we recalled the more or less standard ``Boltzmannian" derivation of macroscopic laws from 
microscopic ones. But since this derivation appeals to probabilistic notions, we tried to relate those notions with 
the issue of what constitutes a valid explanation in the natural sciences.

We claim that one explains a macroscopic behavior on the basis of microscopic laws if, given a macrostate $F_0$, the overwhelming majority of microstates corresponding to $F_0$ give rise, through their deterministic evolution, to an
induced evolution $F_0 \to F_t$ in accordance with the macroscopic law.

We also tried to clarify the status of our ``ignorance" in these explanations.
It is true that we  ignore the initial conditions of the microstates of our system or the details of 
their time evolution, but what we argued in this paper is that
this ignorance does not prevent us from understanding why the system tends towards equilibrium, again because
this is result of the deterministic evolution of the overwhelming majority of the microstates.


\begin{thebibliography}{99}

\bibitem{Ar}  V. I. Arnol'd,  {\it Mathematical Methods of Classical Mechanics}, 
Springer-Verlag New York Inc., 2nd ed., 1989

 \bibitem{BCV}   M. Baldovin, L. Caprini, A. Vulpiani,  Irreversibility and typicality: A simple analytical result
  for the Ehrenfest model, {\it Physica A}, {\bf 524}, 422--429, 2019
  
\bibitem{Bow} R. Bowen, {\it Equilibrium states and the ergodic theory of Anosov diffeomorphisms}, Lecture
Notes in Mathematics, Vol. 470. Springer-Verlag, Berlin-New York, 1975.

\bibitem{BR} R. Bowen, D. Ruelle, The ergodic theory of Axiom A flows, {\it Invent. Math.}, {\bf 29},  181--202, 1975
 
\bibitem{Bri} J. Bricmont, Science of chaos, or chaos in science?
 {\it Physicalia Magazine} {\bf 17}, 159--208, 1995
 
 \bibitem{CCCV}  L. Cerino, F. Cecconi, M. Cencini, A. Vulpiani, The role of the number of degrees of freedom and chaos in
   macroscopic irreversibility,  {\it Physica A}, {\bf 442}, 486--497, 2016
 
 
\bibitem{dBP} S. de Bi\`evre, P.E. Parris, A rigourous demonstration of the validity of Boltzmann's scenario for the spatial homogenization of a freely expanding gas and the equilibration of the Kac ring, {\it  J. Stat. Phys.}, {\bf 168}, 772--793, 2017

 
 
\bibitem{dF1} B. de Finetti, La pr\'evision: ses lois logiques, ses sources subjectives, {\it Annales de l'Institut Henri Poincar\'e}, {\bf 7}, 1--68, 1937

\bibitem{dF2} B. de Finetti, {\it Theory of Probability: A Critical Introductory Treatment},  Wiley, New York, 2017

\bibitem{ER} J-P. Eckmann, D. Ruelle, Ergodic theory of chaos and strange attractors, {\it Rev. Mod. Phys.} {\bf 57}, 617--656, 1985

\bibitem{GO} G. A. Gottwald, M. Oliver, Boltzmann's Dilemma: An Introduction to Statistical Mechanics via the Kac Ring, {\it SIAM Rev.}, {\bf 51}, 613--635, 2009 

 
\bibitem{He1} C. Hempel, The function of general laws in history, {\it Journal of Philosophy}, {\bf 39}, 35--48, 1942,.

\bibitem{He2} C. Hempel,  P. Oppenheim,  Studies in the logic of explanation, {\it Philosophy of Science}, {\bf 15}, 135--175, 1948.

\bibitem{Ja1} E.T. Jaynes, {\it Papers on Probability, Statistics and Statistical Physics}, ed. by R.
D. Rosencrantz, Reidel, Dordrecht, 1983

\bibitem{Ja2} E.T. Jaynes, {\it Probability Theory: the Logic of Science}, Cambridge University Press, Cambridge, 2003 


\bibitem{Ka} M. Kac, {\it Probability and Related Topics in the Physical
Sciences}, Interscience Pub., New York, 1959

\bibitem{La} P.S. Laplace, {\it A Philosophical Essay on Probabilities}, Transl. by F. W. Truscott
and F. L. Emory, Dover Pub., New York, 1951. Original: {\it Essai philosophique
sur les probabilit\'es}, C. Bourgeois, Paris 1986, text of the fifth edition, 1825.

\bibitem{Ma} T. Maudlin, {\it  The Metaphysics Within Physics}, Oxford University Press, Oxford, 2007

\bibitem{Penr} R. Penrose:  {\it The Emperor's New Mind}, Oxford University Press, Oxford, 1989

\bibitem{Ru1} D. Ruelle, A measure associated with axiom-A attractors, {\it  Amer. J. Math.} {\bf 98},  619--654, 1976.

\bibitem{Ru2} D. Ruelle  {\it Thermodynamic Formalism}, Encyclopedia of Mathematics and Its
Applications No 5 Addison Wesley, New York, 1978

\bibitem{Sh} C. E. Shannon,  A mathematical theory of communication, {\it Bell System Technical Journal} {\bf 27}, 379--423, 
 1948
 
 \bibitem{Sin} J. G. Sinai, Gibbs measures in ergodic theory
{\it Russian Mathematical Surveys} {\bf 27}, 21-69, 1972

\bibitem{Tho} C. J. Thompson,
{\it Mathematical Statistical Mechanics}, Princeton University Press,
Princeton, 1972



\end{thebibliography}
\end{document}